\documentclass[cup6a]{cupbook}

\usepackage{graphicx}

\usepackage{epsfig}
\usepackage{graphicx}
\usepackage{float}
\usepackage{color}
\usepackage{txfonts}
\usepackage[authoryear]{natbib}
\usepackage{rotate}
\usepackage{epsf}
\title[ \LaTeXe\ Style Guide for Authors]
      {CUP Standard Designs}
\author{Cambridge \TeX-to-type}
\date{9th July 1997}

\newcommand{\ddt}{\partial_t}
\newcommand{\ddx}{\partial_x}
\newcommand{\ddu}{\partial_u}
\newcommand{\f}{\frac}
\newcommand{\mbfv}{\mathbf{v}}

\newcommand{\mbfB}{\mathbf{B}}
\newcommand{\mbfJ}{\mathbf{J}}
\newcommand{\del}{\mathbf{\nabla}}
\newcommand{\lomega}{\lambda_\Omega}

\newcommand{\Mdot}{\,\dot{\cal M}}
\newcommand{\MMdot}{\hspace{1.5ex}$\tilde{\rule[0.0ex]{0ex}{2ex}}\hspace{-1.9ex}\Mdot$}
\newcommand{\beq}{\begin{equation}}
\newcommand{\eeq}{\end{equation}}
\newcommand{\ben}{\begin{enumerate}}
\newcommand{\een}{\end{enumerate}}
\newcommand{\bit}{\begin{itemize}}
\newcommand{\eit}{\end{itemize}}
\newcommand{\bec}{\begin{center}}
\newcommand{\enc}{\end{center}}
\newcommand{\D}{\partial}
\newcommand{\DD}{\frac}
\newcommand{\ber}{\begin{array}}
\newcommand{\eer}{\end{array}}

\newcommand{\mm }{\mathrm}

\begin{document}
\pagenumbering{roman} \cleardoublepage \pagenumbering{arabic}
\bec Hujeirat, ~A., Keil, B.W. \\

{\small ZAH - Center for Astronomy, Landessternwarte-K\"onigstuhl,
 69117 Heidelberg, Germany}\\
${}$\\
and\\
${}$\\
Heitsch, F.\\

 \small{Department of Astronomy, 500 Church St,
   Ann Arbor, MI 48109-1042, USA}
 \enc

\maketitle \tableofcontents
 \chapter{Advanced numerical methods in astrophysical fluid dynamics}

Computational gas dynamics has become a prominent research field
both in astrophysics and cosmology.  In the first part of this
review we intend to briefly describe several of the numerical
methods used in this field, discuss their range of application and
present strategies for converting conditionally-stable numerical
methods into unconditionally-stable solution procedures.  The
underlying aim of the conversion is to enhance the robustness and
unification of numerical methods and subsequently enlarge their
range of applications considerably. In the second part Fabian
Heitsch presents and discusses the implementation of a time-explicit
MHD Boltzmann solver.

\section{Numerical methods in AFD}
Astrophysical fluid dynamics  (AFD) deals with the properties of
gaseous-matter under a wide variety of circumstances.  Most
astrophysical fluid flows evolve over a large variety of different
time and length scales, henceforth making their analytical treatment
unfeasible.

On the other hand, numerical treatments by means of computer codes
has witnessed an exponential growth during the last two decades due
to the rapid development of hardware technology. Nowadays, the vast
majority of numerical codes are capable of treating large and
sophisticated multi-scale fluid problems  with high resolutions and
even in three-dimensions.

\begin{figure}[htb]
\centering {
\includegraphics*[width=8 cm, bb=1 1 562 534,clip]{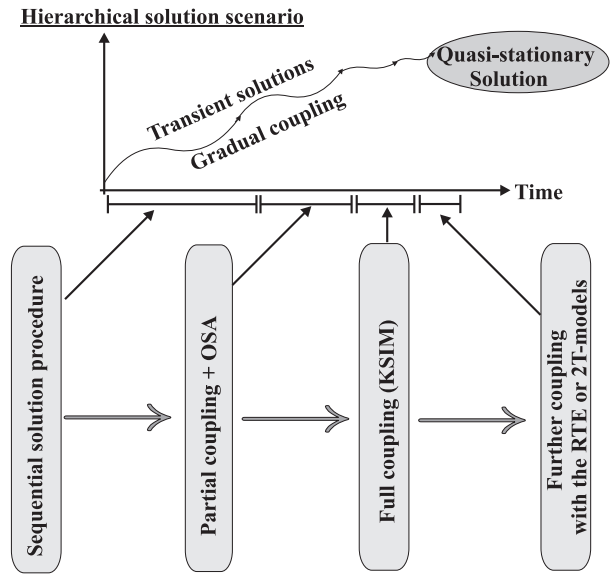}
} \caption{\label{HSS} \small A schematic description of the
hierarchical solution scenario (HSS). The HSS is based on
dynamical-varying the efficiency and robustness of the numerical
method to leapfrog the transient phase. The method is most suitable
for searching quasi-stationary flow-configurations that depend
weakly on the initial conditions. Here the coupling between the
equations can be enhanced gradually, by starting solving them
sequentially, then partial-coupling in combination with the operator
splitting approach (OSA), full-coupling using the Krylov-subspace
iterative method (KSIM) and finally extending the coupling to
include the radiative transfer equation (RTE) and energy equation of
multi-temperature plasmas.}\label{HHS}
\end{figure}

The numerical methods employed in AFD can be classified into two
categories:
\begin{enumerate}
  \item Microscopic oriented methods mostly based on
  N-body (NB), Monte-Carlo (MC) and on the Smoothed Particle Hydrodynamics (SPH).
  \item Grid oriented methods. To this category belong the finite
  difference (FDM), finite volume (FVM) and finite-element methods (FEM).
\end{enumerate}
Most numerical methods used in AFD are conditionally-stable. Hence,
they may converge if   the Courant-Friedlichs-Levy condition for
stability is fulfilled. As long as efficiency is concerned, these
methods are unrivalled candidates for flows that are strongly
time-dependent and compressible. They may stagnate however, if
important physical effects are to be considered or even if the flow
is weakly incompressible. On the other hand, only a small number of
the numerical methods employed in AFD are unconditionally stable.
These are implicit methods, but they are effort-demanding from the
programming point of view.\\
It has been shown that strongly implicit (henceforth IM) and
explicit (henceforth EM) methods are different variants of the same
algebraic problem \citep{Hujeirat2005}. Hence both methods can be
unified within the context of the hierarchical solution scenario
(henceforth HSS, see Fig. \ref{HSS}).

In Table \ref{Table1} we have summarized the relevant properties of
several numerical methods available.
\begin{table}
\begin{minipage}{\linewidth}
\begin{tabular}{l|c|c|c}
   & Explicit & Implicit & HSS  \\\hline
\parbox[c][6ex][c]{0.15 \hsize}{solution\\ method}
   & $q^{n+1} = q^n + \delta t\,d^n$
   &  $q^{n+1} = q^n + \delta t \tilde{A}^{-1}d^*$
   &  {\scriptsize $q^{n+1}  = \alpha q^n + (1-\alpha) \delta t \tilde{A}_d^{-1} d^*$} \\ \hline
Type of flows
   & \parbox{0.23 \hsize}{Strongly time-\\dependent,\\ compressible,\\
weakly dissipative\\ HD and MHD\\ in 1, 2 and 3 dimensions}
   & \parbox{0.23 \hsize}{Stationary, \\quasi-stationary,\\
highly dissipative,\\ radiative and\\ axi-symmetric MHD-flows in 1,
2 and 3 dimensions}
   & \parbox[c][23ex][c]{0.25 \hsize}{Stationary,\\
quasi-stationary,\\ weakly compressible,\\ highly dissipative,\\
radiative and\\ axi-symmetric MHD-flows in 1, 2 and 3 dimensions} \\
\hline
Stability
   & conditioned
   & unconditioned
   & unconditioned \\ \hline
Efficiency
   & $1$ (normalized/2D)
   & $\sim m^2$
   & $\sim m_d^2$ \\ \hline
\parbox[c][9ex][c]{0.15 \hsize}{Efficiency:\\ Enhancement\\ strategies}
   & Parallelization
   & \parbox{0.23 \hsize}{Parallelization,\\ preconditioning,\\ multigrid}
   & \parbox{0.23 \hsize}{HSS, parallelization, preconditioning, prolongation} \\ \hline
\parbox{0.15 \hsize}{Robustness:\\ Enhancement\\ strategies}
   & \parbox[c][12ex][c]{0.23 \hsize}{i. subtime-stepping \\ ii. stiff terms\\ are solved\\ semi-implicitly}
   & \parbox{0.23 \hsize}{i. multiple iteration\\ ii. reducing the time step size}
   & \parbox{0.25 \hsize}{i. multiple iteration\\ ii. reducing the time step size, HSS} \\ \hline
\parbox{0.15 \hsize}{Numerical Codes\\ Newtonian}
   & \parbox[c][12ex][c]{0.23 \hsize}{Solvers1${}^{a}$\\
      {\scriptsize
       ZEUS\&ATHENA${}^{b}$,\\
       FLASH${}^{c}$,
       NIRVANA${}^{d}$,\\
       PLUTO${}^{e}$,
       VAC${}^{f}$ }
      }
   &  Solver2${}^{g}$
   &  IRMHD${}^{h}$ \\ \hline
\parbox{0.15 \hsize}{Numerical Codes\\ Relativistic}
   & \parbox[c][15ex][c]{0.23 \hsize}{Solvers3${}^{i}$\\
      {\scriptsize
       GRMHD${}^{j}$,
       ENZO${}^{k}$,\\
       PLUTO${}^{l}$,
       HARM${}^{m}$,\\
       RAISHIN${}^{n}$,
       RAM${}^{o}$,\\
       GENESIS${}^{p}$,
       WHISKY${}^{q}$ } }
   & Solver4${}^{r}$
   & GR-I-RMHD${}^{s}$ \\ \hline
\end{tabular} \\
  \caption{\label{Table1}
  A list of only a part of the grid-oriented codes in AFD and their algorithmic properties.
  In these equations,  $q^{n,n+1}$,
  ${\delta t}$, ${\tilde{A}}$, $\alpha$ and $d^*$ denote the vector of variables from the old and new
  time levels, time step size, a preconditioning matrix, a switch on/off parameter and a time-modified
  defect vector, respectively. ``m" in row 4 denotes the bandwidth of the corresponding matrix.}
  \smallskip
  \hrule
  \parbox{\hsize}{\footnotesize
  ${}^{a}$\citet{Bodenheimer_etal1978, Clarke1996},
  ${}^{b}$\citet{StoneNorman1992, GardinerStone2006},
  ${}^{c}$\citet{Fryxell_etal2000},
  ${}^{d}$\citet{Ziegler1998},
  ${}^{e}$\citet{MignoneBodo2003, Mignone_etal2007},
  ${}^{f}$\citet{Toth_etal1998},
  ${}^{g}$\citet{Wuchterl1990, Swesty1995},
  ${}^{h}$\citet{Hujeirat1995, Hujeirat2005, Falle2003},
  ${}^{i}$\citet{Koide_etal1999, Komissarov2004},
  ${}^{j}$\citet{DeVilliersHawley2003},
  ${}^{k}$\citet{Oshea_etal2004},
  ${}^{l}$\citet{Mignone_etal2007},
  ${}^{m}$\citet{Gammie_etal2003},
  ${}^{n}$\citet{Mizuno_etal2006},
  ${}^{o}$\citet{ZhangMacFadyen2006},
  ${}^{p}$\citet{Aloy1999},
  ${}^{q}$\citet{Baiotti_etal2003},
  ${}^{r}$\citet{Liebensdoerfer_etal2002},
  ${}^{s}$\citet{Hujeirat2007}. }
\end{minipage}
\end{table}
\begin{figure}[htb]
\centering {
\includegraphics*[width=8 cm, bb=8 1 563 522,clip]{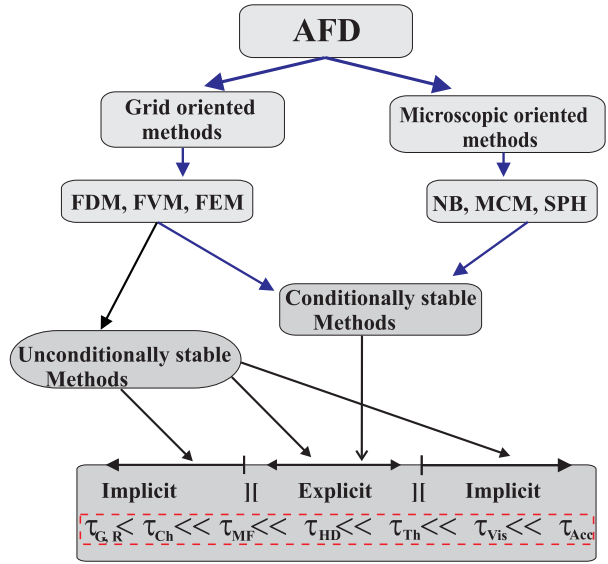}
} \caption{\label{TimeScales} \small Numerical methods (finite
difference, finite volume, finite element, N-Body, Monte Carlo and
the smoothed particle hydrodynamics employed in AFD and their
possible regime of application from the time scale point of view.
The time scales read as follows: the
$\textrm{radiative-}\tau_\mathrm{R},~$
$\textrm{gravitative-}\tau_\mathrm{G},~$
$\textrm{chemical-}\tau_\mathrm{Ch},~$
$\textrm{magnetic-}\tau_\mathrm{MF},~$
$\textrm{hydrodynamic-}\tau_\mathrm{HD},~$
$\textrm{thermal-}\tau_\mathrm{Th},~$
$\textrm{viscous-}\tau_\mathrm{Vis},~$ and the $\textrm{accretion
time scale-}\tau_\mathrm{Acc}.~$
 }
\end{figure}

\section{Time scales in AFD}

Assume we are given a box of $L\times L\times L$ dimensions filled
with a rotating multi-component gaseous-matter. The gas is said to
be  radiating, magnetized, chemical-reacting, partially ionized and
under the influence of its own/external gravitational field. Let the
initial state of the gas be characterized by a constant velocity,
density, temperature and a constant magnetic field. The time-scales
associated with the flow can be obtained directly from the radiative
MHD-equations as follows \citep[see][for detailed description of the
set of equations]{Hujeirat2005}.

\begin{figure}[htb]
\includegraphics*[width=8 cm, bb=1 13 566 592,clip]{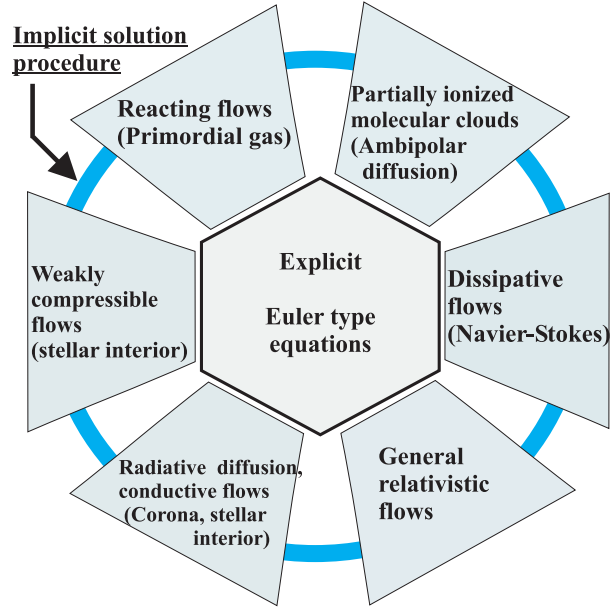}
 \caption{\label{Implicit_Explicit} \small The regime of
application of explicit method is severely limited to Euler-type
flows, whereas sophisticated  treatment of most flow-problems in AFD
require the employment of much more robust methods. }
\end{figure}
%
\begin{table}[htb]
\begin{tabular}{ll|l|l|l}
Scaling&variables &  Molecular cloud & Accretion(onto SMBH)&
Accretion (onto UCO)
\\\hline
$\tilde{L}$   &Length                  & $\mathcal{O}(pc) $
                                           & $\mathcal{O}(\mm{AU})$
                                               &$\mathcal{O}(10^{6,}~\mm{cm})$ \\
$\tilde{\rho}$&Density                 & $10^{-22}~\mm{g~cm}^{-3}$
                                           & $10^{-6}~\mm{g~cm}^{-3}$
                                               & $10^{-8}\rm{~g~cm^{-3}}$\\
$\tilde{\mathcal T}$ &Temperature      & 10 K
                                           & $10^6$ K
                                              & $10^7$ K\\
$\tilde{V}$ &Velocity                  & $0.3~\rm{km~s}^{-1}$
                                           & $10^2~\rm{km~s}^{-1}$
                                              & $10^{2-3}~\rm{km~s}^{-1}$\\
$\tilde{B}$ &Magnetic Fields           & $30~\mu$ G
                                           & $10^2~ \mm{G}$
                                              & $10^4~ \mm{G}$  \\
$\tilde{\mathcal M}$ &Mass             & $10^3~\mm{M}_\odot$
                                           & $10^6~\mm{M}_\odot$
                                              & $~\mm{M}_\odot$\\
\MMdot & Accretion rate        &
                                            &$10^{-2}~\mm{M}_\odot~\mm{Y}^{-1}$
                                               & $10^{-10}~\mm{M}_\odot~\mm{Y}^{-1}$
 \end{tabular}
\caption{A list of possible scaling variables typical for three
different astrophysical phenomena: giant molecular clouds, accretion
onto supermassive black holes (SMBHs) and accretion onto
ultra-compact objects (UCO). These variables may be used for
reformulating the radiative MHD equations in non-dimensional
form.}\label{Define_TSacles}
\end{table}


\bit
\item Continuity equation:
\beq
   \DD{\partial \rho}{\partial t} +   \nabla \cdot \rho V = 0,
\eeq where $\rho,V$ stand for the density and the velocity field.
 Using scaling variables (e.g. Table \ref{Define_TSacles}), we may approximate the terms of
 this equation as follows:
 \[ \DD{\partial \rho}{\partial t} \sim
 \DD{{\rho}}{\tau}~~ \textrm{ and}~~ \nabla \cdot \rho V
 \sim \DD{{\rho}~{V}}{{L}}. \\
 \textrm{ This yields the hydrodynamical time scale}~~ \tau_{HD}=
 \DD{{L}}{{V}}.\\
 \]
The so-called accretion time scale can be obtained by integrating
the continuity equation over the whole fluid volume. Specifically,
\[
\int_{Vol}{\DD{\partial\rho}{\partial t}}\,dVol =\DD{\partial
M}{\partial t}\sim \DD{{M}}{\tau}, ~~~~~~~~~~ \int_{Vol}{(\nabla
\cdot \rho V)}\, d Vol = \int_{S}\rho V \cdot n\cdot dS = \Delta
\dot{M} \sim \dot{M},\\\] where ``Vol" denotes the total volume of
the gas and ``S'' corresponds to its surface. Equating the latter
two terms, we obtain:
\[ \DD{M}{\tau} \sim \dot{M} \Rightarrow \tau_{acc} \sim
\DD{M}{\dot{M}}
 .\]
 In general   $\tau_{acc}$ is one of the longest time scales characterizing
 astrophysical flows connected to the accretion phenomena.

\item The momentum  equations:
\beq
     \DD{\partial V}{\partial t} +  \nabla  V\otimes V
    = -\DD{1}{\rho}\nabla P + f_{cent}  +  \DD{f_\mm{rad}}{\rho} + \nabla \psi
      + \DD{\nabla\times B\times B}{4\pi\rho}  + Q^\mm{r}_\mm{vis},
\eeq where $P,~f_{cent},~ f_\mm{rad},~ \psi,~B,~Q^\mm{r}_\mm{vis}$
denote gas pressure, centrifugal force, radiative force,
gravitational potential, magnetic field and viscous operators,
respectively. From this equation, we may obtain the following time
scales:
\begin{enumerate}
  \item The sound speed crossing time can be obtained by comparing
  the following two terms:
  \[
  \DD{\partial V}{\partial t} \approx \DD{\nabla P}{\rho}, ~~~~\textrm{~~~~which yields:}~~~~
  \tau_{s}\approx \tau_{HD}~\left(\DD{V}{V_s}\right)^2,
  \]
  where $V_s$is the sound speed.
  \item The gravitational time scale:
  \[ \DD{\partial V}{\partial t} \approx \nabla \psi
  \Rightarrow \tau_{G}=~\tau_{HD}~\left(\DD{V}{V_g}\right)^2,\]
  where $V^2_g = GM/L$ and G is the gravitational constant.
  \item Similarly, the Alf$\grave{\rm v}$en-wave crossing-time:
\[ \DD{\partial V}{\partial t} \approx \DD{\nabla\times B\times B}{4\pi\rho}
  \Rightarrow \tau_{mag}=~\tau_{HD}~\left(\DD{V}{V_A}\right)^2,\]
  where $V^2_A (= B^2/4 \pi \rho)$ denotes the Alf$\grave{\rm v}$en speed squared.
\item Radiative effects in moving flows propagate on the radiative scale, which is
      obtained from:
\[ \DD{\D V}{\D t}
\approx \DD{f_\mm{rad}}{\rho}
 \Rightarrow \tau_{rad}=~\tau_{HD}~\left(\DD{V}{c}\right)^2,
 \]
  where c is the speed of light.
  \item The viscous time scale:
\[ \DD{\partial V}{\partial t} \approx Q^\mm{r}_\mm{vis}
\sim \DD{\nu V}{L^2}
  \Rightarrow \tau_{vis}=\DD{L^2}{\nu}\]
  where $\nu$ is a viscosity coefficient.
\end{enumerate}
\item
The induction equation, taking into account the effects of
$\alpha_\mm{dyn}-$dynamo, magnetic diffusivity $\nu_\mm{diff}$ and
of ambipolar diffusion reads:
 \beq
  \DD{\partial B}{\partial t} =  \nabla \times \langle V \times B  + \alpha_\mm{dyn} B
  -\nu_\mm{mag} \nabla \times B\rangle + \nabla\times\{\DD{B}{4 \pi \gamma \rho_i \rho_n}
  \times [B\times(\nabla \times B)]\}, \eeq
  where $\rho_\mm{i,n}$ denote the ion and neutral densities.\\

Thus, the induction equation contains several important time scales:
\begin{enumerate}
  \item The dynamo amplification time scale, which results from the
  equality: \[ \DD{\partial B}{\partial t} = \nabla \times \alpha_\mm{dyn} B \Rightarrow
  \tau_{dyn} = \DD{L}{\alpha_\mm{dyn}}\]
  \item The magnetic-diffusion time scale:
\[ \DD{\partial B}{\partial t} = \nabla \times (\nu_\mm{mag} \nabla \times
B) \Rightarrow  \tau_{diff} = \DD{L^2}{\nu_{mag}}\]
  \item The ambipolar diffusion time scale:
\[ \DD{\partial B}{\partial t} = \nabla\times\{\DD{B}{4 \pi \gamma \rho_i \rho_n}
  \times [B\times(\nabla \times B)]\}\]
  \[\Leftrightarrow \DD{B}{\tau}\sim
  \DD{1}{L}\left(\DD{B^2}{4\pi\rho_n}\right)\left(\DD{1}{\gamma\rho_i}\right)\left(\DD{B}{L}\right)\sim
  \DD{V^2_A}{\gamma\rho_i}\DD{B}{L^2}= \mathcal{D}_{amb} \DD{B}{L^2}
 \Rightarrow \tau_{amb} = \DD{L^2}{\mathcal{D}_{amb}},\]
where $ \mathcal{D}_{amb} (= V^2_A/(\gamma\rho_i)) $ is the
ambipolar diffusion coefficient.
\end{enumerate}
\item The chemical reaction equations.\\
The equation describing the chemical-evolution of  species $``i"$ is
:
\beq \DD{\D\rho_i}{\partial t} = \sum_m \sum_n k_{mn} {\rho_m}
{\rho_n} + \sum_m I_m{\rho_m},\eeq
 where $k_{mn}$ denotes the
reaction rate between the species ${m}$ and ${n}.$ $I_m$ stands for
other external sources. For example, the reaction equation of atomic
hydrogen in a primordial gas reads:
\[ \DD{\D\rho_{H}}{\partial t} = \DD{k_2}{m_H} \rho_{H^+}\rho_e - \DD{k_1}{m_H}\rho_H\rho_e
\Leftrightarrow \DD{\rho_{H}}{\tau} \sim
\DD{k_2}{m_H}~\rho_H~\rho_e\Rightarrow \tau_{ch}\sim
\DD{m_H}{k_2~\rho_e},
\]
where $\rho_e,~k_2(10^{-10}~\rm cm^{3}~s^{-1})$ correspond to the
electron density and to the generation rate of atomic hydrogen
through the capture of electrons by ionized atomic hydrogen. $~m_H$
corresponds to the mass of atomic hydrogen.

\item Equations of relativistic MHD

The velocities in relativistic  flows are comparable to the speed of
light. This implies that the hydrodynamical  $\tau_{HD}$ and
radiative $\tau_{rad}$ time scales are comparable and that both are
much shorter than in Newtonian flows. \eit

\begin{table}[htb]
\begin{tabular}{l|l|l|l}
Time scales &  Molecular cloud & Accretion(onto SMBH)& Accretion
(onto UCO)
\\\hline
$\tau_{HD}$    & $\sim 10^6 \rm{~Yr}$
                  & $\sim$ months
                     & $\sim 1$ s  \\
$\tau_{rad}/\tau_{HD}$   & $\sim 10^{-6}$
                             & $\sim 10^{-3}$
                                &$\sim 10^{-3}$ \\
$\tau_{grav}/\tau_{HD}$  & $\sim 10^{-2}$
                            & $\sim 10^{-3}$
                                & $\sim 10^{-3}$\\
$\tau_{ch}/\tau_{HD}$    & $\sim 10^{-1}$
                              & $\sim 10^{-5}$
                                   & $\sim 10^{-4}$ \\
$\tau_{mag}/\tau_{HD}$   & $\sim 10^{-2}$
                               & $\sim 10^{0}$
                                     & $\sim 10^{-1}$\\
$\tau_{vis}/\tau_{HD}$   & $\sim 10^{1}$
                             & $\sim 10^{2}$
                                  & $\sim 10^{2}$\\
$\tau_{acc}/\tau_{HD}$   &
                               &  $\sim 10^{4}$
                                      & $\sim 10^{12}$ \\
 \end{tabular}
\caption{A list of the time scales relative to the hydrodynamical
time scale for three different astrophysical
phenomena.}\label{TScale_nach_Thd}
\end{table}
We note that although the dynamical time scale in relativistically
moving flows is relatively short, there are still several reasons
that justify the use of implicit numerical procedures. In
particular:
 \ben
\item The relativistic MHD equations are strongly non-linear,
       giving rise to fast growing non-linear perturbations,
        imposing thereby a further restriction on the size of the time step.
\item    The deformation
      of the geometry grows non-linearly when approaching the black hole. Thus, in order to
       capture flow-configurations  in  the vicinity of a black hole
       accurately, a non-linear distribution of the grid points is
       necessary, which, again, may destabilize explicit schemes.
\item  Initially non-relativistic flows may become ultra-relativistic or vice versa.
       However, almost all non-relativistic astrophysical flows known to date are considered to be
        dissipative and diffusive.
       Therefore, in order to track their time-evolution reliably, the  employed numerical solver
       should be capable of treating  the corresponding second order viscous terms properly.
\item  The accumulated round off errors resulting from
       performing a large number of time-extrapolations for
       time-advancing a numerical hydrodynamical solution may easily
       cause divergence. The constraining effects of boundary
       conditions may fail to configure the final numerical solution.
\een

\section{Numerical methods: a unification approach}\label{UnifiedScheme}

In this section we show that explicit and implicit methods are
special cases of a more general solution method in higher
dimensions.\\
 Assume we are given the following evolution equation of a vector variable $q$:
 \beq
\DD{\D q}{\D t} + L(q) = f,\label{unieq1}
 \eeq
where $L,f$ correspond to an advection operator and to external forces.\\

Adopting a time-forward discretization  procedure, the unknown
vector $q$ at the new time level can be extrapolated as follows:

\beq q^{n+1} =  q^{n} + \delta t \cdot RHS^{n}, \label{unieq2}
 \eeq
where $RHS= f - L(q).$

Depending on the time step size and on the number of grid points,
the
numerical procedure can be made sufficiently accurate in space and time.\\
On the other hand Equation \ref{unieq2} can be viewed as an equality
of two one-dimensional vectors:
\beq \textrm{[vector of unknowns] = [vector of knowns] }
\Leftrightarrow~~~~~~ q^{n+1} = \bar{b}, \label{unieq3}\eeq
where $\bar{b} = q^{n} + \delta t \cdot RHS^{n}.$ \\
In higher dimensions, however, Equation \ref{unieq3} is a special
case of the matrix equation: \beq A q^{n+1} =
\bar{b}\label{2deq},\eeq in which it is projected along the diagonal
elements. It is obvious that the matrix ${I/\delta t}$ is a further
simplification of
 the  matrix that contains just the diagonal elements of A.

Therefore, we may adopt the higher dimension formulation to gain a
better understanding of the stability of the solution procedure.
\begin{figure}[htb]
\centering {
\includegraphics*[width=8cm, bb=52 47 400 300,clip]{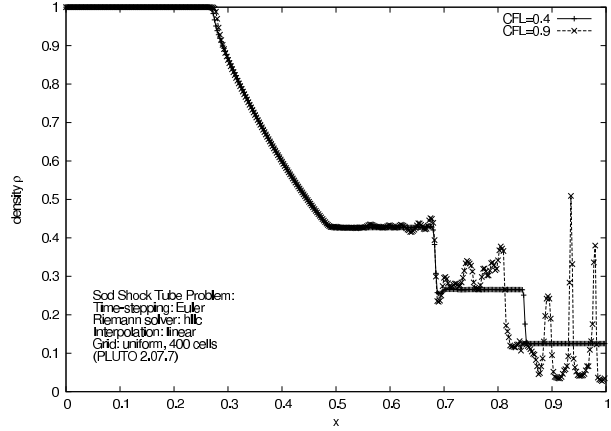}
} \caption{\label{STP_PLUTO} \small The the profile of the shock
tube problem obtained with Courant-Friedrichs-Levy numbers CFL=0.4
and 0.9 using the PLUTO code. Although both CFL-numbers are smaller
than unity the numerical solution procedure does not appear to be
stable even with CFL=0.9.}\label{SodPluto}
\end{figure}
According to matrix algebra, a necessary condition for the matrix A
to have a stable inversion procedure is that A must be strictly
diagonally dominant. Equivalently,  the entries in each row of the
matrix A must fulfill the following condition: the module of the
diagonal element $d_{i,i}$ is larger than the sum of all
off-diagonal elements \(\sum_\mm{j\neq i}{|a_{i,j}|},\) where i and
j denote the row and column numbers of the matrix . Applying a
conservative and monotonicity preserving scheme, the latter
inequality may be re-written in the following form:
\beq \left|\DD{1}{\delta t} + \textrm{positive contributions}\right|
> \sum_\mm{j(\neq i)}{|a_{i,j}|}. \eeq
We note that since $\delta t$ is a free parameter, it can be chosen
sufficiently small, so that $1/\delta t$ largely dominates all other
off-diagonal elements, or so large that $1/\delta t$ becomes
negligibly small.

We may further simplify this inequality by choosing the time step
size even smaller, such that
 \beq \DD{1}{\delta t}
> \sum_j |a_{i,j}|,~~~~~ \textrm{for $\forall~~j\neq i$} \eeq can be safely
fulfilled. We may decompose the matrix A as follows:
\[A = D + R = D(I+D^{-1}R),\] where D is the matrix consisting of the diagonal entries
of A and $R~(= A-D)$ consists of the off-diagonals.  Thus, the
elements of D are proportional to $1/\delta t$, whereas those of R
are proportional to $\delta t$. This implies that A can be expanded
around $I/\delta t $ in the form: \beq A = A^{(0)} + A^{(1)} +
A^{(2)} + \cdot\cdot\cdot,\eeq
 where the leading matrix
 $A^{(0)}\approx \DD{1}{\delta \rm t}~\rm I~$ and $A^{(1)} \sim \delta \rm t~I.$
 In this case the inversion
of the matrix A is not more necessary and the resulting numerical
procedure would correspond to a classical time-explicit method.\\

\subsection{Example}\label{example}
The time-evolution of density in one-dimension   is described by the
continuity equation: \beq L_\rho = \DD{\D \rho}{\D t} + \DD{\D \rho
U}{\D x} = 0.\eeq The corresponding Jacobian matrix is: $A =
\D{L_\rho}/{\D \rho}.$ The non-zero entries of A read: \beq a_{ii}=
\DD{1}{\delta t} + \DD{|U_i|}{\Delta x}\textrm{~~~~~and the off
diagonal~~~} a_{ij}= -\DD{|U_{j+1}|}{\Delta x} \textrm{~~~~for
i$\neq$ j~~~~,   }\eeq where ${\Delta x}$ and $i,j$ denote the grid
spacing and grid point numbering. Applying a first order upwind
discretization, then the condition of diagonal dominance demands:
\beq \left|\DD{1}{\delta t} + \DD{U_i}{\Delta x}\right| >
\DD{|U_{j+1}|}{\Delta x}.\eeq This condition can be further
simplified by choosing the time step size so small, such that
\beq \DD{1}{\delta t} > \DD{2~\max\,(|U_{j}|, |U_{j+1}|)}{\Delta x}
\Leftrightarrow  \DD{{\delta t} \,\max\,(|U_{j}|, |U_{j+1}|)}{\Delta
x} < \DD{1}{2}.\eeq

Thus, the condition of diagonal dominance is more restrictive than
the normal CFL condition. This may explain, why most explicit
methods fail to converge for Courant-Friedrichs-\-Levy number CFL $=
1 -\epsilon~~~$ (see Fig. \ref{SodPluto}).

\begin{figure}[htb]
\centering {\hspace*{-0.5cm}
\includegraphics*[width=13 cm, bb=0 0 490 140,clip]{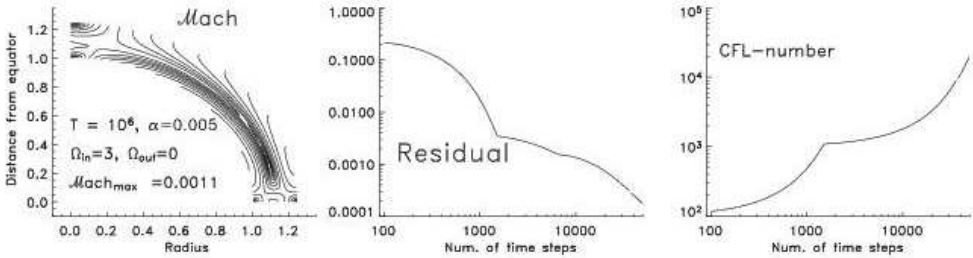}
} \caption{\label{WInko3} \small  Weakly incompressible flow between
two concentric rotating spheres. Left panel: the 2D-distribution of
$\textrm{Mach number (= V/V}_\mathrm{s})~~$ is displayed (25
isolines) for extreme weakly incompressible flows $(Max~(Mach)\sim~
10^{-3}).$ The maximum residual (middle) and the CFL-number (right)
versus the number of time steps are shown. }\label{Incompress}
\end{figure}
\section{Converting time-explicit into implicit solution
methods}\label{Num_Convert}
 In a series of publications, we have shown that the
robustness of explicit methods can be enhanced gradually to recover
full-implicit solution procedures \citep[see][]{Hujeirat2005}. In
the following we outline the main algorithmic steps towards
extending classical explicit methods into implicit:
\begin{enumerate}
  \item Use the same mathematical form of $RHS^n$ of  Eq. \ref{unieq2} to compute $RHS^{n+1}= RHS(q^{n+1})$
         and subsequently the mean $\overline{RHS}= \alpha\cdot{RHS^n} +
         (1-\alpha)\cdot RHS^{n+1},$ where $0 \leq \alpha \leq 1$ is a parameter
         that may depend also on the time step size.
  \item  Define the defect \beq d= -(\DD{q^{n+1}-q^n}{\delta t}) + \overline{RHS}. \eeq
  \item  Compute the Jacobian $J^{real}= \D L_q/\D q$, where $L_q$
  denotes the set of equations in operator form.
  \item Construct a  simplified matrix $\tilde{A}$ (preconditioner), which is easy
  to invert, but still share the spectral properties of $J^{real}$
  \citep{Hackbusch}.

\begin{figure}[htb]
\centering {
\includegraphics*[width=6 cm, bb=0 0 405 325,clip]{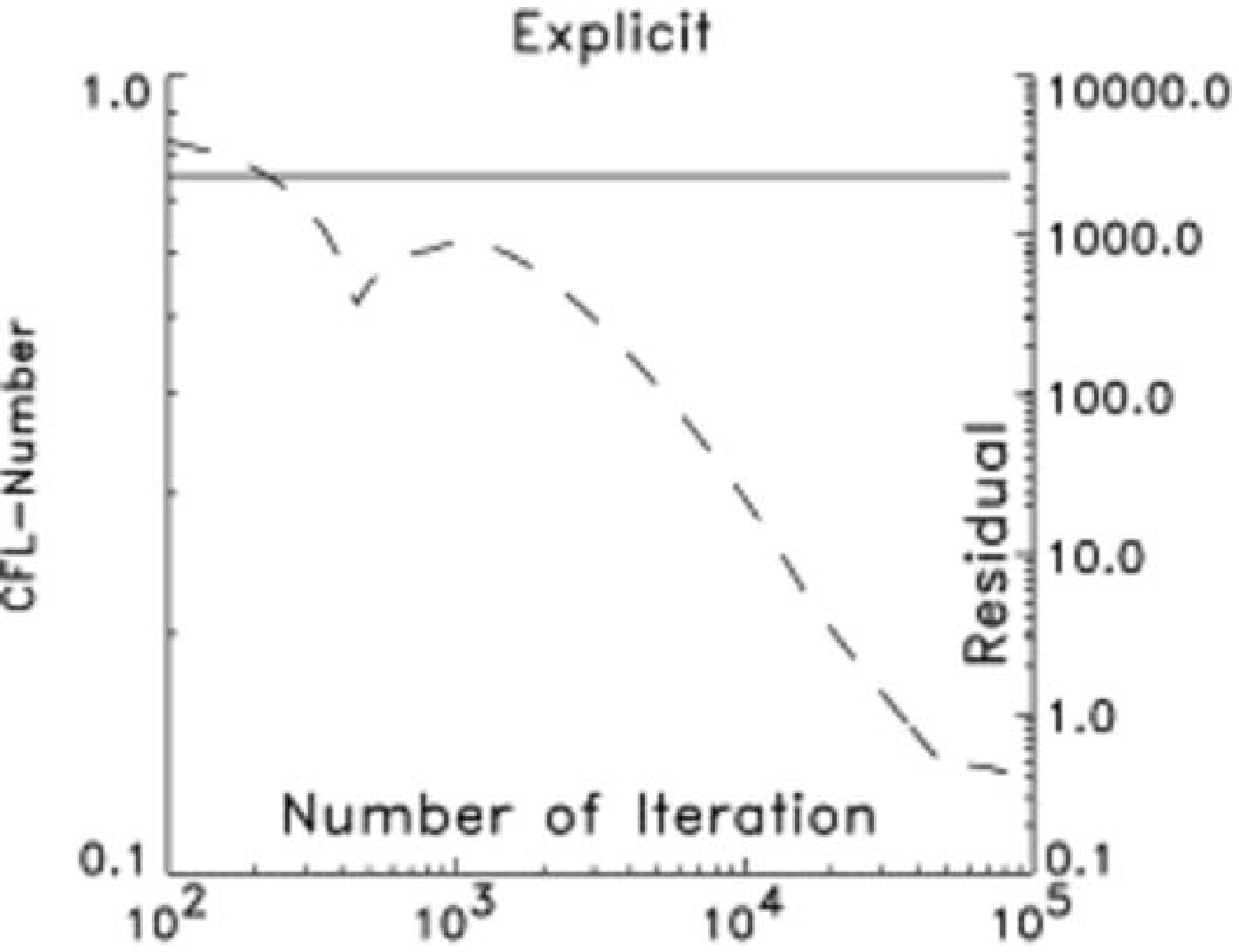}
\includegraphics*[width=6 cm, bb=0 0 405 325,clip]{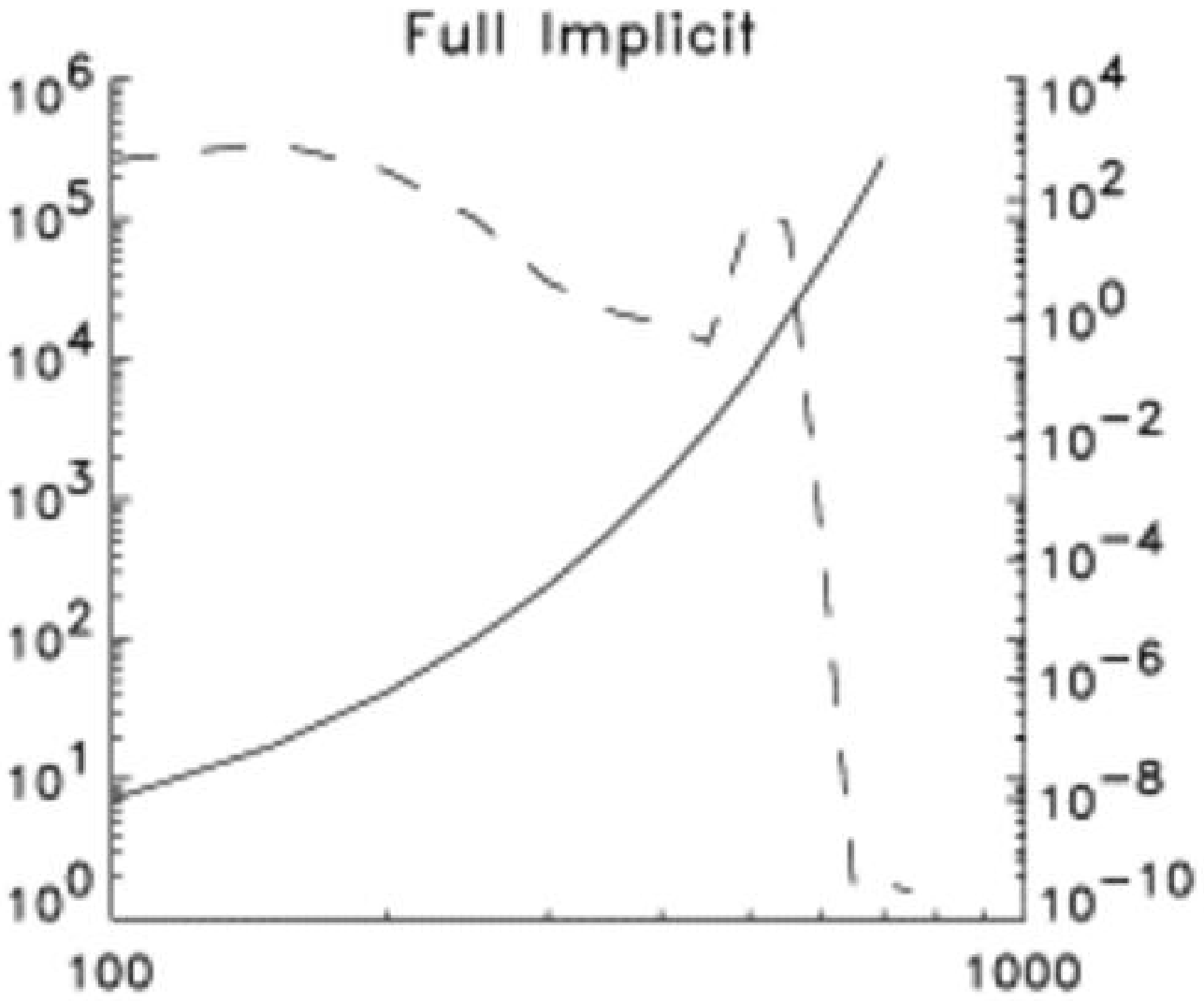}
} \caption{\label{WInko3} \small The profiles of the CFL-number
(solid line) versus the number of iteration both for explicit and
implicit solution procedures (dashed line). The profiles correspond
to the free-fall of spherical plasma onto a non-magnetized
Schwarzschild black hole, in which the final solution is
time-independent. }\label{FFall}
\end{figure}
  \item Solve the system of equation: \beq  \tilde{A} \mu = d, \label{conveq2}\eeq
  where $\mu$ is a vector of small correction, so that $q^{l+1}=q^{l}+ \mu.
  $\\
  In general $\tilde{A} \neq J^{real}, $ which implies that Equation
  \ref{conveq2}
  should be solved iteratively to assure that the maximum norm of the defect, $||d||_\infty,$ is sufficiently small.
\end{enumerate}

We note that for sufficiently small $\delta t$, the matrix $I/\delta
t$ can be made similar to  $J^{real}$, hence they share the same
spectral space. As a consequence, a variety of
   solution procedures can be constructed that range from purely explicit up to
   strongly implicit, depending on how similar  the preconditioner $\tilde{A}$ is to
   the real Jacobian. This naturally suggests  the hierarchical solution scenario
    as a highly powerful numerical algorithm
   for enhancing the robustness of explicit schemes and optimizing
   their efficiency \citep[Fig. \ref{HSS}, see also][]{ Hujeirat2005}


\section{Summary-I}\label{summary}
In this part of the review we have presented a method for converting
conditionally-stable explicit methods into numerically stable
implicit solution procedures. The conversion method allows a
considerable enlargement of the range of application of explicit
methods. The hierarchical solution scenario is best suited for
gradual enhancement of their robustness and optimizing their
efficiency.

\clearpage

\newpage
\bec Part II\\

{(Magneto-)Hydrodynamic Boltzmann Solvers} \enc

In this part, I will discuss the implementation of a time-explicit
gas-kinetic grid-based integrator for non-relativistic hydrodynamics
introduced by Prendergast \& Xu (1993), Xu (1999) and Tang \& Xu
(2000), and its extension to non-ideal magneto-hydrodynamics
(Heitsch et al. 2004, 2007). Some properties of Boltzmann solvers
are discussed in \S\ref{s:whyboltzmann}, the equations and the
implementation are described in \S\ref{s:proteus}, followed by a
selection of test cases and applications (\S\ref{s:testcases}) and a
summary (\S\ref{s:summary}).

\section{Why Boltzmann Solvers?\label{s:whyboltzmann}}

It is the physical model for the fluid equations which distinguishes
gas-kinetic schemes from the widely popular Godunov methods. The
latter are formulated on the basis of the Vlasov-equation, i.e.
assuming that any dynamical time scale is larger than the collision
time between particles, setting the collision term in the Boltzmann
equation to zero. The distribution function is then given by a
Maxwellian at all times. In contrast, gas-kinetic schemes keep the
collision term in the Boltzmann equation, but because of the
impractibility to compute all the collisions between particles, they
need to come up with a model for the collision term.

One such model has been introduced by Bhatnagar, Gross \& Krook
(1954), formulating the collision term as the difference between the
equilibrium distribution function $g$ (the Maxwellian) and the
initial distribution function $f$, resulting in a Boltzmann equation
of the form
\begin{equation}
  \ddt f + u \ddx f + \dot{u}\ddu f = \frac{g-f}\tau\label{e:boltzmann},
\end{equation}
where $\tau$ is the collision time. Integrating
eq.~\ref{e:boltzmann} over a time $t$ gives (at position $x$)
\begin{equation}
  f(x,t,u) = \f{1}{\tau}\int_0^tg(x-u(t-t'),t',u)\,e^{-(t-t')/\tau}
dt'+e^{-t/\tau}\,f_0(x-ut,0,u),
\end{equation}
where $\tau$ is the collision time, and $f_0$ the initial
distribution function. For a complete description, see Xu (2001).
Thus, the distribution function $f$ at time $t$ gets two
contributions: one from the decaying initial conditions $f(t=0)$,
and one from the growing equilibrium distribution $g$.

The 0th, 1st and 2nd order velocity moments of the distribution
function (here for a monatomic gas)
\begin{equation}
g \equiv
\rho\left(\frac{\lambda}{\pi}\right)^{3/2}\,\exp(\lambda(\mathbf{u}-\mathbf{U})^2)
\end{equation}
result in the (macroscopic) conserved quantities density $\rho$,
momentum density $\rho U$ and total energy density $\rho E$. The
quantity $\lambda\equiv m/(2kT)$. The corresponding moments of the
Boltzmann equation~\ref{e:boltzmann} give the conservation
equations. The BGK collision term in eq.~\ref{e:boltzmann} gives
then rise to a viscous flux, depending on the ratio of the CFL time
step and a specified collision time. Thus, the Reynolds number of
the flow can be controlled. The Prandtl number is 1 by construction.
The scheme is upwind and it satisfies the entropy condition
(Prendergast \& Xu 1993, Xu 2001). The fully controlled dissipative
term come at (close) to no extra computational cost. Fragmentation
of hydrodynamically unstable systems due to numerical noise thus can
be suppressed. Specifically, gas-kinetic schemes can easily provide
a viscosity independent of grid geometry, thus allowing e.g. the
modeling of disks on a cartesian grid (see Slyz et al. 2002).

In the following I will discuss a specific implementation of a
gas-kinetic solver, namely Proteus (see Heitsch et al. 2007).

\section{Equations and Implementation: Proteus\label{s:proteus}}

Proteus solves the equations of non-ideal magnetohydrodynamics, with
an Ohmic resistivity $\lomega$, and a shear viscosity $\nu$.

\begin{equation}\label{e:continuity}
\ddt\rho + \del\cdot (\rho\mbfv) = 0
\end{equation}
\begin{equation}\label{e:momentum}
\ddt \rho\mbfv + \del\cdot \left[\rho\mbfv\mbfv -
\f{\mbfB\mbfB}{4\pi} + p + \f{\mbfB^{2}}{8\pi}\right] =
\del\cdot\bar{{\bf \Pi}}
\end{equation}
\begin{equation}\label{e:energy}
\ddt\rho E + \del\cdot \left[\rho E \mbfv +
  (p+\f{\mbfB^{2}}{8\pi})\mbfv-\f{(\mbfv\cdot\mbfB)\mbfB}{4\pi}\right]=
 \mbfv\cdot\left(\del\cdot\bar{{\bf \Pi}}\right) + \lomega\mbfJ^{2}
\end{equation}
\begin{equation}\label{e:induction}
\ddt\mbfB + \del\cdot (\mbfv\mbfB - \mbfB\mbfv) =
\lomega\del^{2}\mbfB,
\end{equation}


The mechanism how to split the fluxes at the cell walls is described
in detail by Xu (1999) and will not be repeated here. Viscosity and
resistivity are implemented as dissipative fluxes. They require
spatially constant coefficients $\lambda_\Omega$ and $\nu$.
Ambipolar drift is implemented in the two-fluid description,
currently only for an isothermal equation of state, though.

Higher-order time accuracy is achieved by a TVD Runge-Kutta time
stepping (Shu \& Osher 1988). For second-order spatial accuracy, a
choice of reconstruction prescriptions is available.

Proteus offers two gas-kinetic solvers, the one just described, and
a one-step integrator at 2nd order in time and space for
hydrodynamics. The latter has been discussed in detail by Slyz \&
Prendergast (1999) and Slyz et al. (2005), so that we refer the
interested reader to those papers.

\section{Test Cases and Applications\label{s:testcases}}

\subsection{1D: Resistively damped Linear Alfven Wave}

\begin{figure}
  \centerline{\includegraphics[width=0.7\columnwidth]{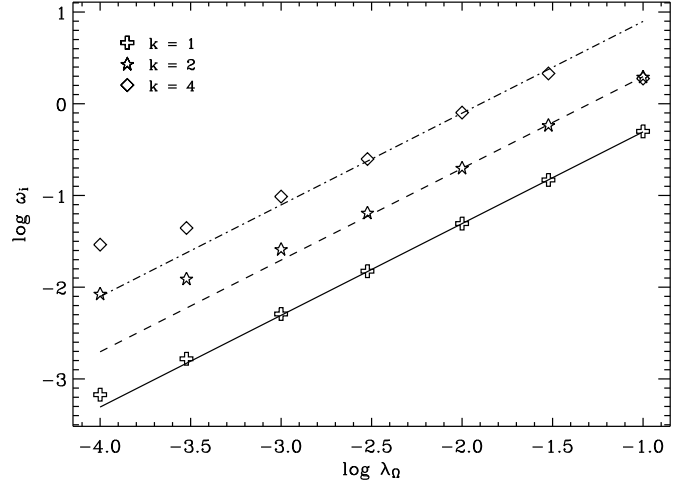}}
  \caption{\label{f:alfven}Logarithm of the damping rate (eq.~[\ref{e:alfven}]) of a
           linear Alfv\'{e}n wave against logarithm of the Ohmic resistivity for
           $\kappa=1,2,4$. The resolution is $N=64$.
           Lines denote the analytical solution.}
\end{figure}

This one-dimensional test checks the resistive flux implementation
as well as the accuracy of teh overall scheme. A linear Alf\'{e}n
wave under weak Ohmic dissipation is damped at a rate of
\begin{equation}
  \omega_i = \frac{1}{2}\lomega k^2\label{e:alfven}
\end{equation}
where $\lomega$ is the Ohmic resistivity, and $k=2\pi\kappa/L$ is
the wave number of the Alfv\'{e}n wave, with $\kappa$ a natural
number. The strongly damped case, where the decay dominates the time
evolution, is uninteresting for our application, since the Ohmic
resistivity is mainly used to control numerical dissipation.
Figure~\ref{f:alfven} shows the damping rate against Ohmic
resistivity $\lomega$ for $\kappa=1,2,4$ at a grid resolution of
$N=64$. The damping rate is derived by measuring the amplitude of
the wave at each full wave period.

From Figure~\ref{f:alfven}, it is clear that, as one diminishes the
value of
 $\lomega$, there comes a point when the numerical resistivity of the scheme
becomes comparable to the physical one, causing the measured damping
rate to flatten out and depart from the analytical solution. For
$\kappa=4$ and $\lomega=0.1$, the wave decays too quickly to allow a
reliable measurement, and the system enters the strongly damped
branch of the dispersion relation. However, we emphasize that even
at $16$ cells per wave length the resistivity range available to
Proteus spans nearly two orders of magnitude.

\subsection{1D: Linear Alfven Waves in Weakly Ionized Plasmas}

The dispersion relation for a linear Alfv\'{e}n wave in a weakly
ionized plasma splits into two branches (Kulsrud \& Pearce 1969): a
strongly coupled branch, for which the ion Alfv\'{e}n frequency
$\omega_k\equiv kB/\sqrt{4\pi\rho_i}\ll \nu_{in}\equiv\gamma
\rho_n$, the ion-neutral collision frequency, and a weakly coupled
branch, for which $\omega_k\gg\nu_{in}\sqrt{\rho_i/\rho_n}$. The
strongly coupled case leads to a dispersion relation of
\begin{equation}
  \omega = \pm\left(\omega_k^2\epsilon-\frac{\omega_k^4}{4\nu_{in}}\right)^{1/2}
  -\imath\frac{\omega_k^2}{2\nu_{in}},
\end{equation}
with $\epsilon\equiv\rho_i/rho_n$. Thus, the strongly coupled
Alfv\'{e}n wave travels at the neutral Alfv\'{e}n speed
$c_{An}\equiv B/\sqrt{4\pi\rho_n}$ and is increasingly damped with
decreasing collision frequency. The weakly coupled branch leads to
\begin{equation}
  \omega=\pm\left(\omega_k^2-\frac{\nu_{in}^2}{4}\right)^{1/2}-\imath\frac{\nu_{in}}{2}.
\end{equation}
Now, the wave travels at the ion Alfv\'{e}n speed, and damping is
proportional to $\nu_{in}$. Since $c_{An}\sqrt{\rho_n/\rho_i}$, the
speeds can be widely disparate.

\begin{figure}
   \centerline{\includegraphics[width=0.7\columnwidth]{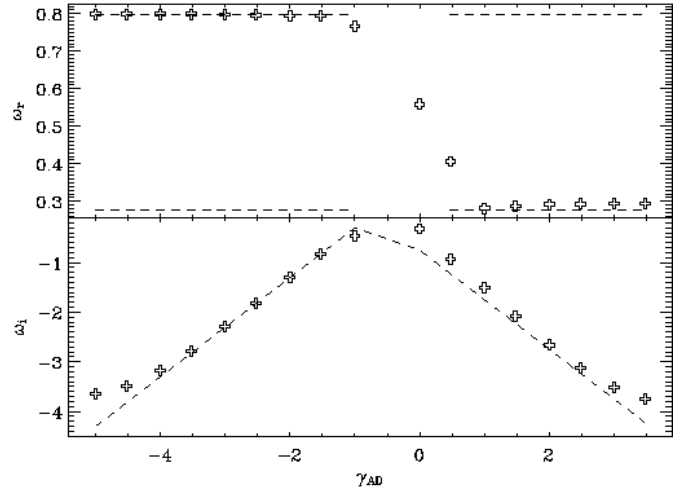}}
  \caption{\label{f:alfvenad}Logarithm of the frequency (upper panel) and damping
rate
           (lower panel) for the linear Alfv\'{e}n wave in a partially ionized plasma.
           For simplicity, we vary the collision coefficient $\gamma_{AD}$ instead of
           the density.}
\end{figure}

Figure~\ref{f:alfvenad} shows the real and imaginary part of the
Alfv\'{e}n wave frequency in a weakly ionized plasma. For
simplicity, we vary the collision coefficient $\gamma_{AD}$ and keep
the densities constant. Wave speed (upper panel) and damping term
(lower panel) are well reproduced.

\subsection{2D: Current Sheet\label{ss:currentsheet}}
\begin{figure}
  \centerline{\includegraphics[width=0.7\columnwidth]{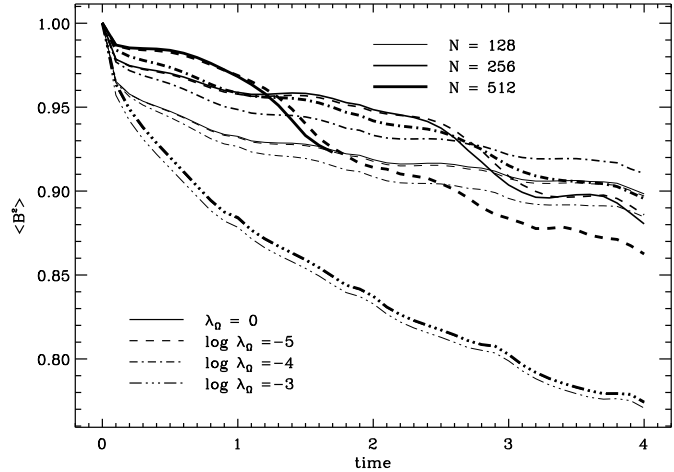}}
  \caption{\label{f:emagtime}Current sheet test. Magnetic energy density $\langle
B^2\rangle$
           against time. A finite resistivity $\lomega$ helps stabilize the code. Line
           thickness stands for resolution, line style for resistivity.}
\end{figure}

This test is taken from Gardiner \& Stone (2005). A square domain of
extent $0\leq x,y\leq 2$ and of constant density $\rho_0=1$ and
pressure $p_0=0.1$ is permeated by a magnetic field along the $y$
direction such that $B_y(0.5<x<1.5)=-1$, and $B_y=1$ elsewhere. The
ratio of thermal over magnetic pressure is $\beta = 0.2$. This setup
results in two magnetic null lines, which then are perturbed by
velocities $v_x=v_0\sin(\pi y)$. Here, we use an adiabatic exponent
of $\gamma=5/3$ and employ the conservative formulation of the
scheme. Figure~\ref{f:emagtime} summarizes the test results in the
form of the  magnetic energy density $\langle B^2\rangle$ against
time. Different line styles stand for resistivities, and the line
thickness denotes the model resolution. We ran tests at $N=128^2$,
$256^2$ and $512^2$. All models ran up to $t=4$ and farther except
for the $512^2$-model at $\lomega=0$. A finite resistivity helps
stabilizing the code.

The evolution of the system follows that described by Gardiner \&
Stone (2005), including the merging of magnetic islands until there
are two islands per magnetic null line left, located approximately
at the velocity anti-nodes. For zero resistivity (solid lines), the
magnetic energy decay depends strongly on the resolution. This
effect is reduced by increasing $\lomega$. For $\log\lomega=-5$
(dashed lines), the energy evolution follows pretty much the curves
for $\lomega=0$ (solid lines), indicating insufficient resolution.
For $\log\lomega=-4$, the two higher resolutions start to separate
from the lower resolution run, while at $\log\lomega=-3$, the two
higher resolutions lead to indistinguishable curves (dash-3dot
lines).

\subsection{2D: Advection of a Field Loop\label{ss:fieldloop}}

\begin{figure}
  \centerline{\includegraphics[width=0.7\columnwidth]{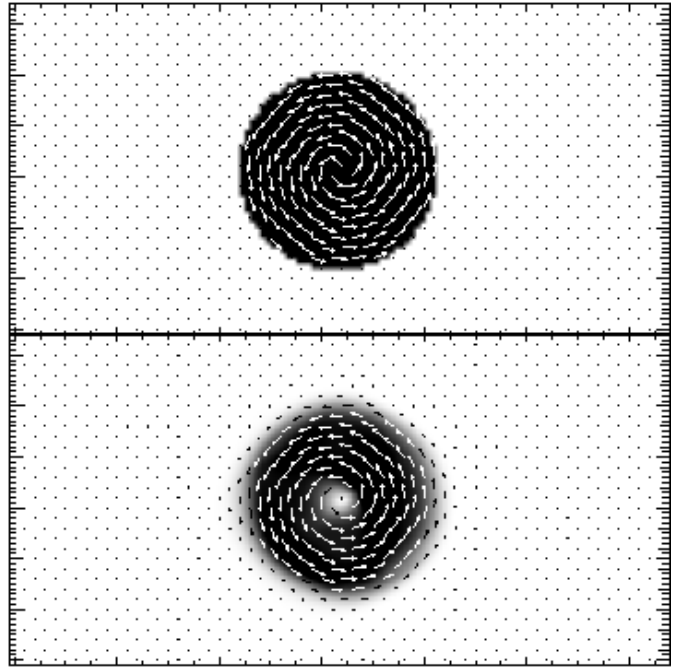}}
  \caption{\label{f:fieldloop}Field loop advection test:
           magnetic energy density $B^2$ at $t=0$ ({\em top}) and
           at $t=2$ corresponding to two horizontal crossing times ({\em bottom}),
           with over-plotted field vectors. The grid resolution is $N_x\times N_y =
128\times 64$.}
\end{figure}

A cylindrical current distribution (i.e. a field loop) is advected
diagonally across the simulation domain. Again, we follow the
implementation presented by Gardiner \& Stone (2005). Density and
pressure are both initially uniform at $\rho_0=1$ and $p_0=1$, and
the fluid is described as an ideal gas with an adiabatic exponent of
$\gamma=5/3$. The computational grid at a resolution of $N_x\times
N_y = 128\times 64$ extends over $-1.0\leq x \leq 1.0$ and $-0.5\leq
y \leq 0.5$. The field loop is initialized via the $z$-component of
the vector potential $A_z = a_0(R-r)$, where $a_0 = 10^{-3}$,
$R=0.3$ and $r\equiv (x^2+y^2)^{1/2}$. The loop is advected at an
angle of $30$ degrees with respect to the $x$-axis. Thus, two round
trips in $x$ correspond to one crossing in $y$.
Figure~\ref{f:fieldloop} shows the initial magnetic energy density
$B^2$ with the magnetic field vectors over-plotted ({\em top}), and
the $B^2$ distribution after two time-units measured in horizontal
crossing times ({\em bottom}). The overall shape is preserved,
although some artifacts are visible. These results concerning the
shape are similar to those of Gardiner \& Stone (2005),
specifically, Proteus preserves the circular field lines. This test
uses $\lomega \equiv 0$.

\begin{figure}
  \centerline{\includegraphics[width=0.7\columnwidth]{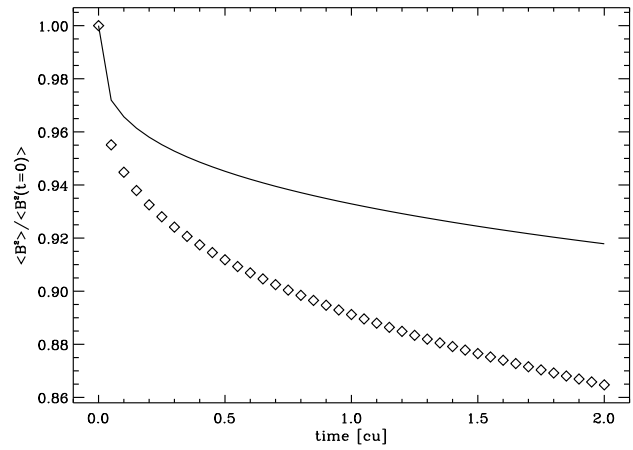}}
  \caption{\label{f:decayloop}Normalized magnetic energy density against time
  (in units of horizontal crossing time) with the same parameters as in
Figure~\ref{f:fieldloop}.
  Diamonds stand for Proteus results, and the energy evolution as observed in ATHENA
  is shown by the solid line.}
\end{figure}

The time evolution of the magnetic energy density corresponding to
Figure~\ref{f:fieldloop} is shown in Figure~\ref{f:decayloop}.
Diamonds stand for Proteus results, the energy decay observed by
Gardiner \& Stone (2005) with ATHENA is indicated by the solid line,
following their analytical fit. The energies are normalized to 1.
Clearly, Proteus is somewhat more diffusive.

In summary, these numerical test cases demonstrate that Proteus
models dissipative MHD effects accurately. Furthermore, it can
advect geometrically complex magnetic field patterns properly.

\section{Summary\label{s:summary}}

Gas-kinetic schemes provide a robust and physical mechanism to solve
the equations of magneto-hydrodynamics. Dissipative effects can be
fully controlled. I discussed a specific implementation of a
gas-kinetic solver -- Proteus --, including resistivity and
(two-fluid) ambipolar diffusion. Details of the implementation have
been presented elsewhere (Tang \& Xu~2000, Heitsch et al. 2004,
2007), and an application to shear flows in magnetized fluids will
be discussed by Palotti et al.~(2008).

\end{document}